\documentclass[]{spie}  

\usepackage{amsmath,amsfonts,amssymb}
\usepackage{graphicx}
\usepackage[colorlinks=true, allcolors=blue]{hyperref}
\usepackage{xspace}

\def\m87{M87$^*$\xspace}

\def\lsim{\mathrel{\raise.3ex\hbox{$<$\kern-.75em\lower1ex\hbox{$\sim$}}}}
\def\gsim{\mathrel{\raise.3ex\hbox{$>$\kern-.75em\lower1ex\hbox{$\sim$}}}}

\title{Receivers for the Black Hole Explorer (BHEX) Mission}

\author[1]{C. Edward Tong}
\author[2]{Kazunori Akiyama}
\author[1]{Paul Grimes}
\author[3]{Mareki Honma}
\author[1]{Janice Houston}
\author[1]{Michael D. Johnson}
\author[4]{Daniel P. Marrone}
\author[1]{Hannah Rana}
\author[5]{Yoshinori Uzawa}

\affil[1]{Center for Astrophysics $|$ Harvard \& Smithsonian, 60 Garden St, Cambridge, MA 02138, USA}
\affil[2]{MIT Haystack Observatory, Westford, MA 01886}
\affil[3]{Mizusawa VLBI Observatory, NAOJ, Iwate, 023-0861, Japan}
\affil[4]{University of Arizona, Steward Observatory, Tucson, AZ 85719, USA}
\affil[5]{Advanced Technology Center, NAOJ, Mitaka, Tokyo, 181-8588, Japan}

\authorinfo{Corresponding Author: C. Edward Tong (\url{etong@cfa.harvard.edu})}

\begin{document} 
\maketitle

\begin{abstract}
In this paper, we introduce the receiver architecture for the Black Hole Explorer (BHEX) Mission, designed to reveal the photon ring of black holes. The primary instrument is a dual-polarization receiver operating over the 240-320 GHz frequency range, utilizing a Superconductor-Insulator-Superconductor (SIS) mixer. This Double-Side-Band (DSB) receiver has an intermediate frequency (IF) range of 4-12 GHz and operates at a bath temperature of 4.5 K, for optimal performance, which necessitates the integration of a cryocooler.
Complementing the primary receiver is a secondary unit covering the 80-106 GHz spectrum, featuring a cryogenic low noise amplifier. This secondary receiver, affixed to the cryocooler’s 20 K stage, serves to augment the SIS receiver’s performance by employing the Frequency Phase Transfer technique to boost the signal-to-noise ratio at the correlator output.
Together, this sophisticated receiver duo is engineered to achieve the quantum-limited sensitivity required to detect the photon ring of black holes, marking a breakthrough in astrophysical observation.

\end{abstract}


\section{Introduction}
\label{sec:intro}  

The Black Hole Explorer (BHEX) is a mission concept designed to study black holes and their jets, and to  observe the sharp photon ring feature from light that has orbited just outside the event horizon \cite{Johnson_2024A}. Working together with a network of ground-based radio telescopes, BHEX will perform space Very-Long-Baseline (VLBI) observations, delivering unprecedented angular resolution. Recent advances in receiver and cryogenic technology are bringing this ambitious goal within reach, with receivers offering quantum-limited sensitivities across a wide frequency range. The planned mission targets a 2025 NASA Announcement of Opportunity for Small Explorer (SMEX) spacecraft.

The main BHEX scientific instrument is a superconducting receiver operating between 240 and 320 GHz. For this receiver to operate correctly, a cryocooler providing a base temperature of 4.5 K is indispensable. The details of this space cryocooler are reported elsewhere \cite{Rana_2024A}. A second receiver, covering 80 to 106 GHz, is based on High Electron Mobility Transistor (HEMT) amplifiers and will be connected to the second stage of the cryocooler for operation at a physical temperature of around 20 K. Using the Frequency-Phase-Transfer (FPT) technique \cite{2015Rioja,2023Rioja}, the lower frequency receiver will be used to enhance the performance of the main science receiver in the VLBI mode. 

Besides delivering the required sensitivities, the receiver design focuses on low power consumption as the resources available on the SMEX spacecraft are limited. Therefore, a number of trade-offs have been made to reduce the resource requirements of the receiver system. In this paper, we will report the current receiver layout and design, with a special emphasis of how our design adapts to the constraints of the spacecraft  while maintaining the high performance required for the mission.

\section{Optical Layout}
\label{sec:optics}
The current BHEX antenna design is a shaped axially symmetric dual reflector system with a 3.5 m aperture and is optimized for maximum aperture efficiency. The details of the antenna design are reported by \cite{BHEX_Sridharan_2024}. One advantage of a shaped antenna design is that the beam feeding the antenna can be made smaller than the beam for a classical Cassegrain design with the same diameter dish, because this design affords a higher edge taper of illumination. As a result, it allows for a very compact optical layout with smaller beam sizes.

\begin{figure}
  \centering    
    \includegraphics[width=5in]{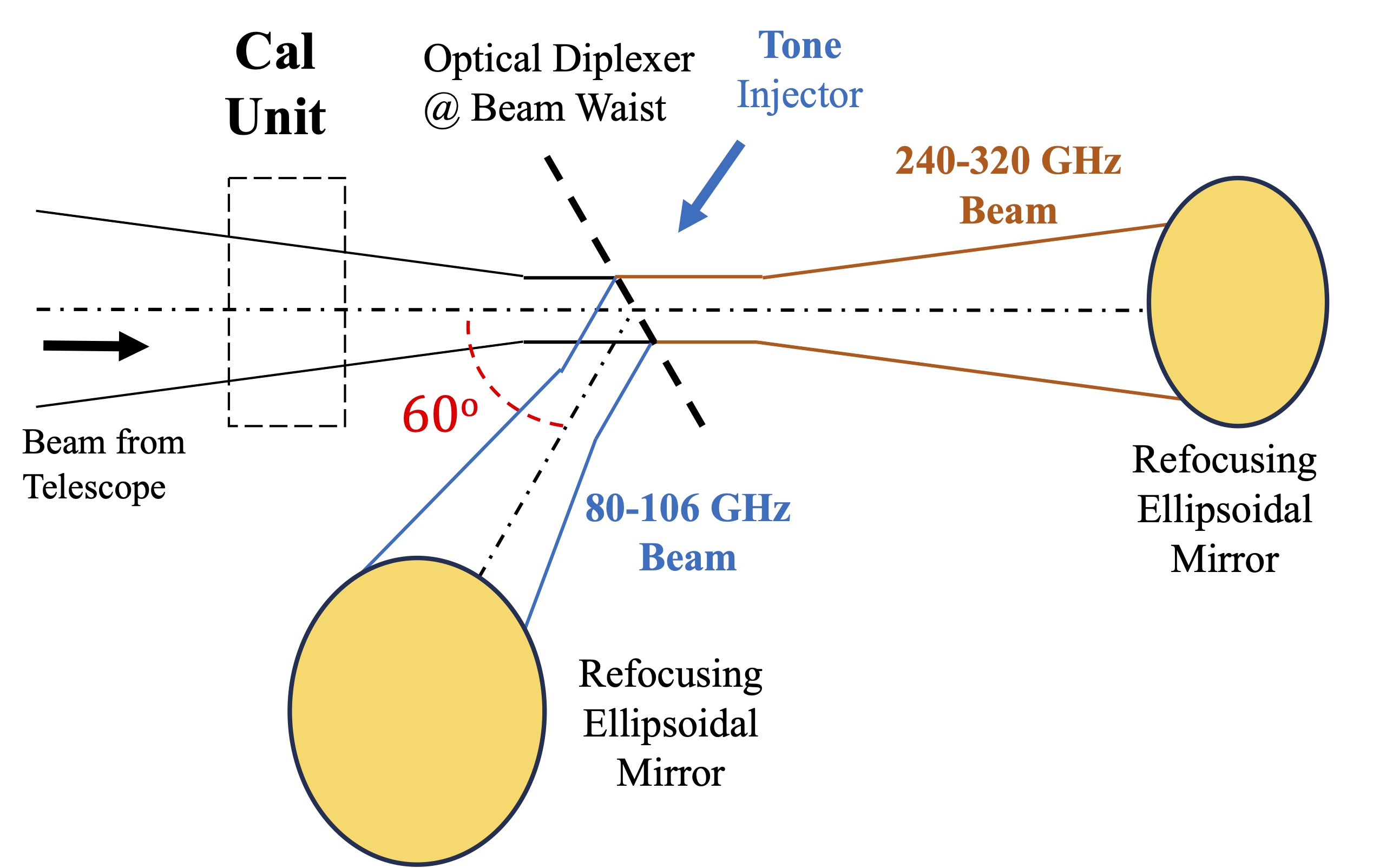}
  \caption{Schematic for BHEX optical layout. The optical diplexer is placed at the secondary focus (beam waist) of the antenna. The low frequency (80-106 GHz) beam is reflected from the diplexer at an angle of 60$^o$ while the high frequency (240-320 GHz) beam is transmitted through the diplexer. Each of the 2 beams is refocused by an ellipsoidal mirror sending the beam away from the plane of incidence towards the 2 receivers.}
  
  \label{fig:optics}
\end{figure}

Referring to Fig~\ref{fig:optics}, the secondary focus of the antenna is located behind the primary antenna. A calibration unit, consisting of a flip mirror and a calibration load, is placed in front of the secondary focus. At the secondary focus, an optical diplexer is installed to separate frequencies above 200 GHz and frequencies around 100 GHz. The design of this diplexer will be based on a dielectric stack \cite{Carter_2024} consisting of disks of high resisitivity silicon alternating with plastic disks. The diplexer is expected to incur a loss of around 3\%. The beams emerging from the diplexer will be re-focused by a focusing mirror before directed towards the horn of their respective receivers.

A tone/comb generator is set up to direct its beam towards the backside of the optical diplexer. This arrangement allows the generator to be weakly coupled to the receiver. When the generator is in operation, the sinusoidal signals will appear in the output of the receiver and the phase coherence of the system can be monitored.

\section{240-320 GHz SIS Receiver}
\label{sec:SIS}

In the initial planning phase of the mission, a dual band receiver set, consisting of a 230 GHz and a 345 GHz receiver, was considered. Observations at 230 GHz are expected to deliver a better signal-to-noise performance for most targets, while those 345 GHz receivers would afford longer baselines, in terms of G$\lambda$, and therefore higher resolution. There is also a larger network of ground based telescopes operating at 230 GHz to support the space VLBI operations. However, given the cost constraints of a SMEX mission, a trade study concluded that a band that covers most of the range between those frequencies (240-320 GHz) could deliver the needed performance while reducing cost.  

The heart of the receiver is a Superconducting-Insulator-Superconducting (SIS) tunnel junction mixer, which is the instrument of choice for the most sensitive receivers operating in this frequency range. This class of receivers offer quantum-limited sensitivities when operated at a physical temperature less than or equal to half the superconducting critical temperature of the SIS junction. For the Niobium-based SIS mixers that are commonly used, this translates into a bath temperature of 4.5 K, which is provided by the space cryocooler. The design goal of the SIS receiver is to provide a 2-photon sensitivity, or a noise temperature 
of 2 $h\nu/k_B$, which translates into Double-Side-Band (DSB) noise temperatures of 23 - 30 K for 240-320 GHz. 

The SIS mixer is a mature technology: SIS mixers based on based on Niobium tunnel junctions with an Aluminum oxide barrier are commonly found in most radio telescopes. We are working with the National Astronomical Observatory of Japan (NAOJ) to develop SIS devices needed for this mission. SIS devices fabricated by NAOJ based on the standard Nb/Al-AlOx/Nb junctions have been used for ALMA band 4, 8, and 10 receivers\cite{Uzawa_2021}. Recently high quality Nb/Al-AlN/Al/Nb junctions with a critical current density of 31kA/cm$^2$ has been used to develop a broadband receiver covering 275-500 GHz \cite{Kojima_2018}.
The baseline design for BHEX SIS receiver requires SIS junctions with a
critical current density of around 12 - 15 kA/cm$^2$, although devices based on Aluminum nitride barrier offering much higher critical current densities may be considered. 

The target RF bandwidth of the receiver is similar to many of the receivers in operation. For example, the wSMA Low Band receiver \cite{Grimes_2020} covering 210 - 270 GHz, for which the ratio of $F_{\rm max}/F_{\rm min} = 1.33$, has the same fractional bandwidth. Therefore, if higher critical current density SIS devices are available, one can consider extending the lower frequency end of the receiver down to 220 GHz. The Local Oscillator (LO) module, described later, is expected to operate down to 220 GHz.

\begin{figure}
    \centering
    \includegraphics[width=4in]{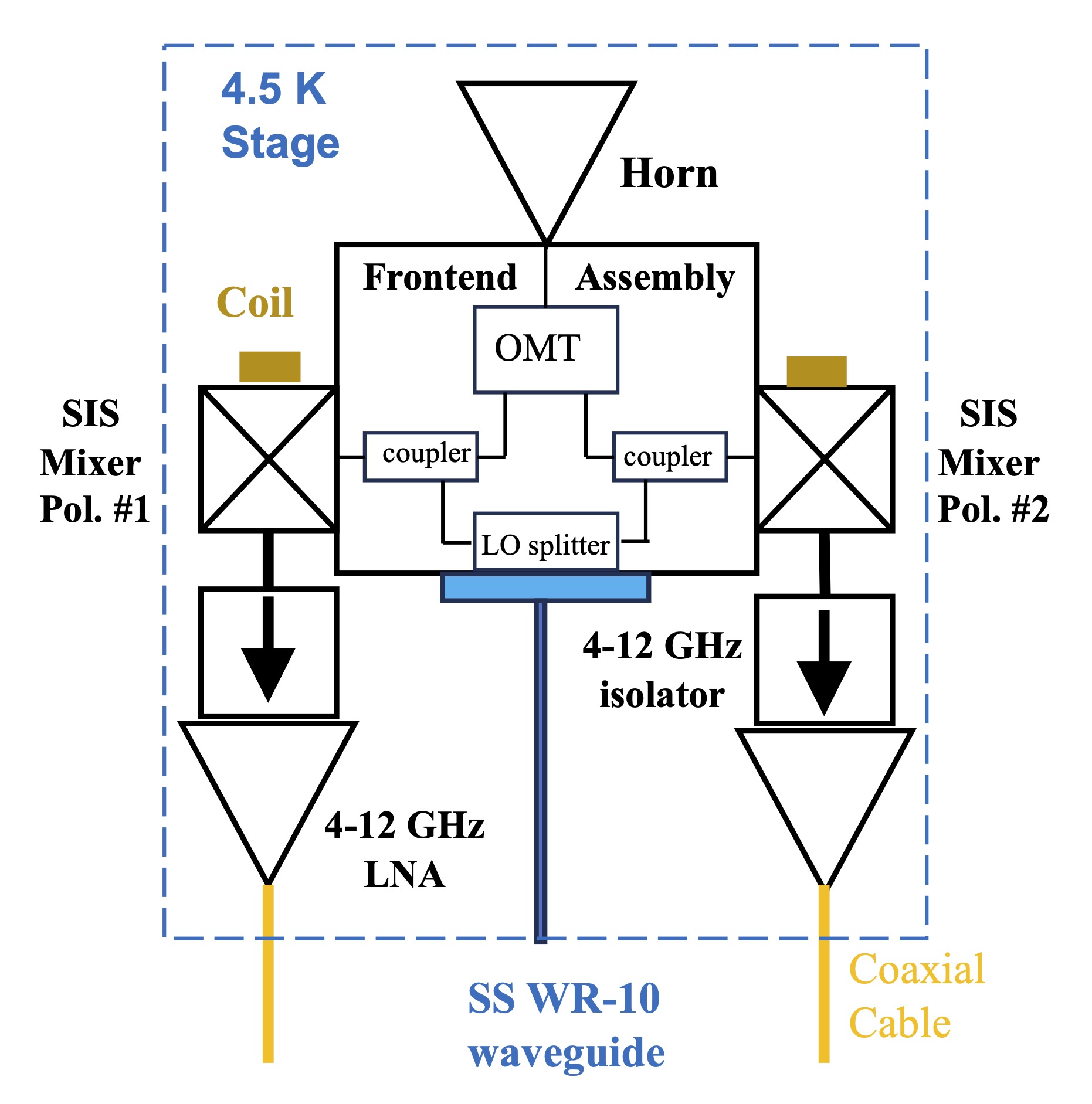}
    \caption{Layout of the SIS receiver anchored to the 4.5 K stage of the cryocooler.}
\label{fig:SIS240}
\end{figure}

The receiver also operates in dual linear polarization. The layout of the receiver is shown in Fig~\ref{fig:SIS240}. A corrugated horn couples the incoming beam into a frontend module, comprising an orthomode transducer, which  separates the two linear polarizations, as well as the LO couplers and a pair of mixer modules, one for each polarization. While many state-of-the-art SIS receivers adopt the sideband separating (2SB) configuration employing a pair of phased-up mixers per polarization, we decided to stay with the more conventional single-ended DSB mixer configuration, which offers a more compact architecture. As a result, both sidebands of the receiver are folded into a single intermediate frequency output. This has the distinct advantage of reducing the output bandwidth of the receiver, halving the required transmission bandwidth of the lasercom system. See further discussion in \cite{BHEX_Marrone_2024}.

The receiver's IF is 4-12 GHz. A 4-12 GHz isolator will be deployed at the output of the SIS mixer, followed by a cryogenic low noise amplifier, both anchored to the 4.5 K stage of the cryocooler. The amplifier pair will operate in a low DC power mode, dissipating 1 mW of power each, while delivering 20 dB gain for the IF. One more stage of amplification will be installed in the 100 K stage of the cryocooler. This setup makes the best use of the limited cooling power available from the space cryocooler. The expected heat load on the 4.5 K stage of the cryocooler is about 10 mW. 

\section{80-106 GHz Receiver}
\label{sec:HEMT}

While the 240-320 GHz SIS receiver is tasked with providing fringes at long baselines to reveal the photon ring of black holes, the 80-106 GHz receiver is needed for the implementation of the Frequency Phase Transfer (FPT) technique\cite{2015Rioja}. Observations at 300 GHz involving long ground - space baselines face many practical difficulties compared to observations at 100 GHz. These include higher atmospheric attenuation, shorter atmospheric coherence time, and higher system noise temperature of ground-based stations, plus a small increase of the noise temperature of the receiver on the spacecraft . By observing simultaneously at both 100 and 300 GHz, the FPT technique permits atmospheric phase corruptions tracked by the low-frequency receiver with a higher signal-to-noise ratio at the correlator output, to be used to calibrate the measurements made by the high-frequency receiver, which has a lower signal-to-noise ratio. The successful implementation of FPT effectively increases the coherent integration time of the main science instrument and thus improves the effective sensitivity of the high-frequency observation.

\begin{figure}
    \centering\includegraphics[width=5in]{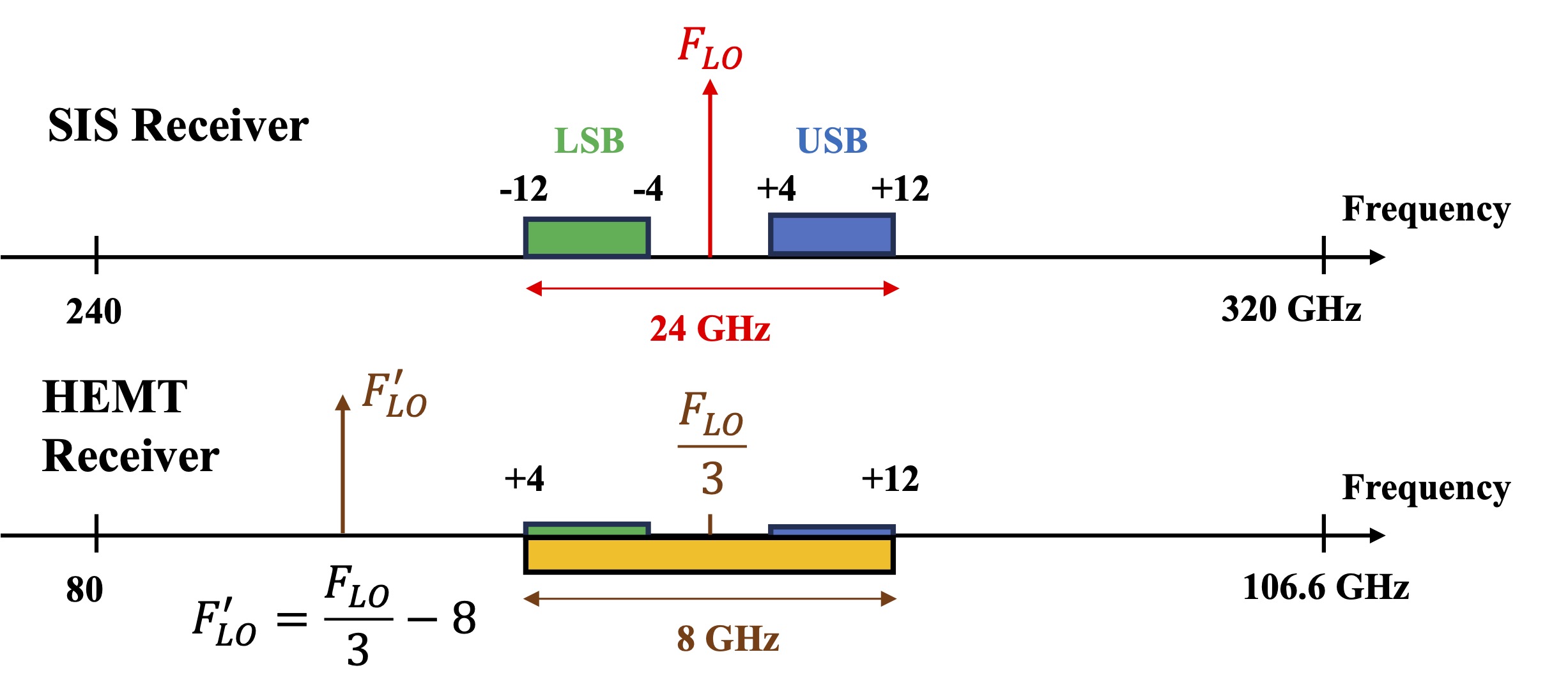}
    \caption{Spectral coverage of the two BHEX receivers. The operating frequency of the HEMT receiver is scaled from that of the SIS receiver by a factor of 3. Both sidebands of the SIS receiver, spanning a width of 24 GHz, correspond to an 8 GHz spectral coverage of for the HEMT receiver. By choosing an appropriate LO frequency, $F'_{LO}$ for the down-converter, together with an appropriate High Pass Filter, the required spectral coverage can be mapped into the USB of the down-converter.}
\label{fig:sscheme}
\end{figure}

In order to simplify the implementation of the FPT techniques, we fix the ratio of the observation frequency at the SIS receiver to that of the 100 GHz receiver to be a ratio of 3. That is why the 240-320 GHz passband of the SIS receiver is mapped to 80-106 GHz for the 100 GHz receiver. As mentioned above, the SIS receiver operates in a DSB mode with a 4-12 GHz IF. Given a sky frequency ratio of 3, the 100 GHz receiver only needs to provide an 8 GHz wide of IF bandwidth to cover the required RF bandwidth. This spectral relationship is illustrated in Fig~\ref{fig:sscheme}.

A commercially available cryogenic HEMT amplifier is chosen to anchor the 80-106 GHz receiver. While this amplifier delivers the lowest noise temperature at 5 K, we choose to operate it at a bath temperature of 20 K to reduce the thermal load at the 4.5 K stage of the cryocooler. Each stage of the amplifier is expected to dissipate 10 mW and delivers a gain of 20-25 dB. The noise penalty of operating at 20 K instead of 4.5 K is estimated to be about 5 K. The expected noise temperature of the receiver is around 45 K.

The beam entering the receiver is coupled to a corrugated feed horn, and a waveguide orthomode transducer (OMT) separates the dual polarization input into 2 outputs. A pair of cryogenic HEMT amplifiers is connected to each output port of the OMT, with a waveguide isolator inserted between the amplifier to improve the passband flatness of the receiver. The output of each polarization channel is directed to a down-converter located at the ambient portion of the instrument through of stainless steel waveguide sections.

\begin{figure}
\centering
    \includegraphics[width=5in]{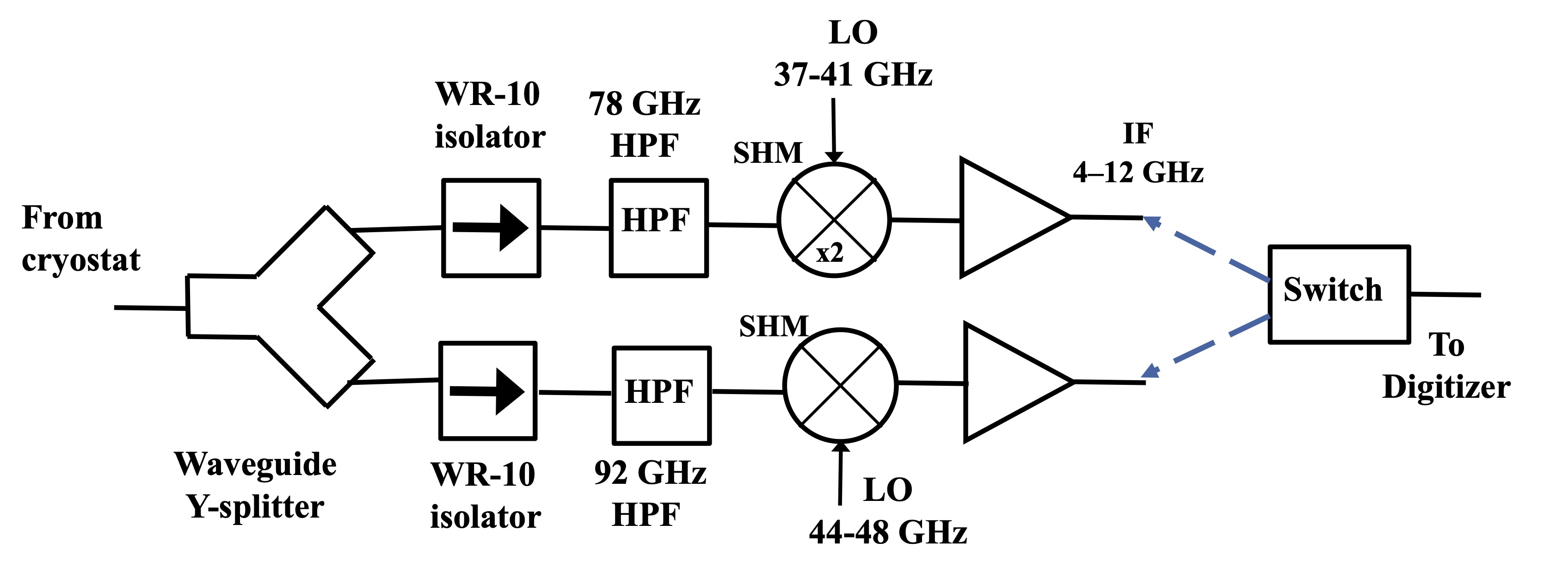}
    \caption{Down Converter for 80-106 GHz receiver. The RF input from the cryostat is divided into 2 streams, one covering the lower frequency portion and the other one the higher frequency portion of the passband. A second harmonic mixer (SHM) down-converts the RF signal into a 4-12 GHz IF. A switch selects the IF signals.}
\label{fig:Dconvert100}
\end{figure}

The layout of the down-converter module is shown in Fig.~\ref{fig:Dconvert100}. Inside each down-converter, the RF is split into 2 streams and a waveguide high pass filter (HPF) is used as a single-side-band filter before the down-conversion with a second harmonic mixer to a 4-12 GHz IF. With this fixed HPF setup, one stream covers the high end of the 86-106 GHz band while the stream covers the lower end of the band. The advantage of the second harmonic mixer is that the required LO runs in the range of 40-53 GHz, which is easier to produce.

\section{Local Oscillator}

\begin{figure}
\centering
\includegraphics[width=5in]{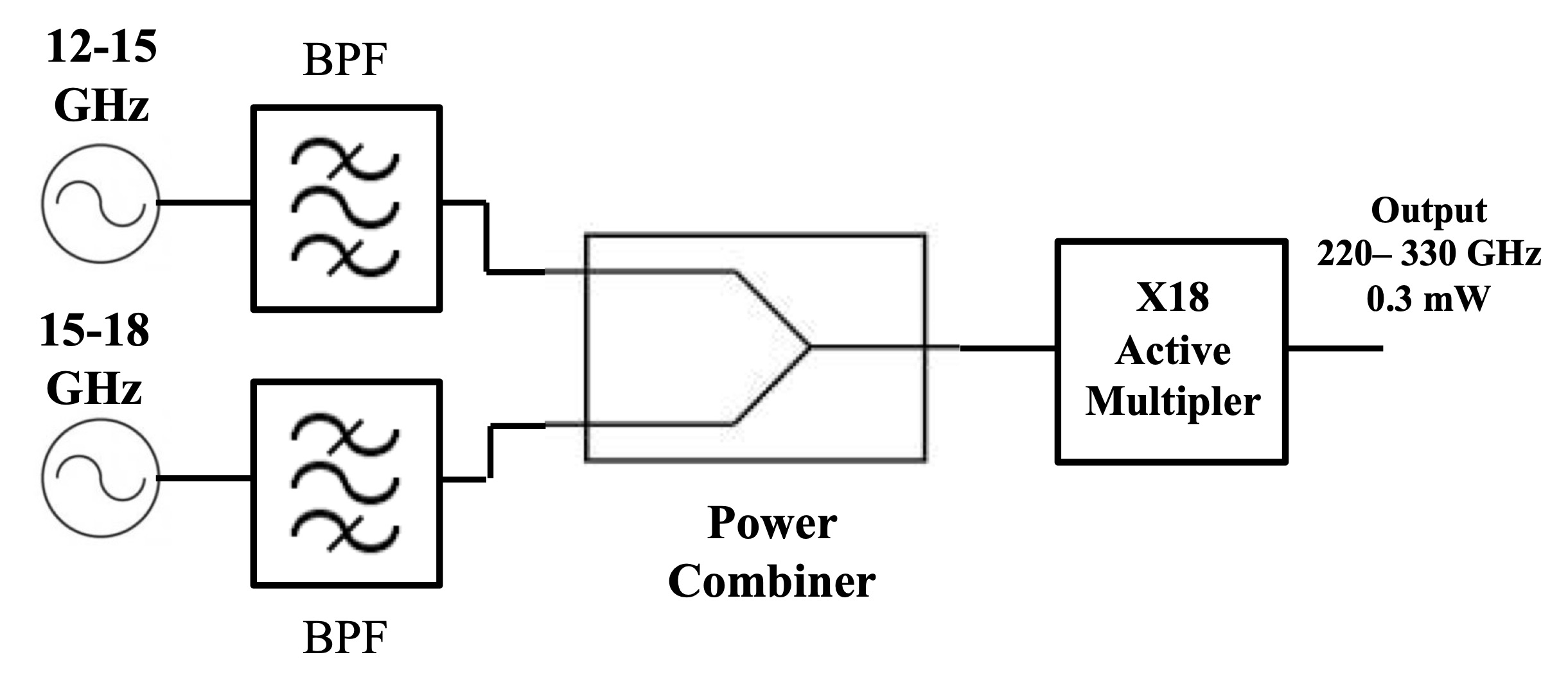}
\caption{Local Oscillator module for BHEX SIS Receiver. A pair of synthesizers are used to cover the required LO frequency range. The required frequency multiplication is 18. The output of the LO module is in a standard WR-3.4 waveguide.}
\label{fig:LO240}
\end{figure}

The SIS receiver will be operated by a Local Oscillator (LO) module located at the ambient part of the spacecraft. A standard WR-3.4 source module, usable between 220 and 330 GHz can cover the required frequency range. That is why if the SIS receiver delivers low noise performance beyond its baseline frequency range of 240-320 GHz, we would be able to extend the frequency coverage to 220 GHz. The LO module will be driven by a synthesizer, which operates at 1/18th of the intended LO frequency. In order to provide redundancy and to limit AM noise, one possibility is to use a pair of synthesizers, one covering the lower portion of the band and the other the higher portion. The two signals will be band-limited and pass through a power combiner to drive the LO module. This arrangement is illustrated in Fig~\ref{fig:LO240}. Only one of the 2 synthesizers will be active during observation. 

The 80-106 GHz HEMT receiver also requires an LO. Since second harmonic mixers are used in the down-converter, the required LO frequency coverage is between 40 and 53 GHz. The LO set up is similar to that of the SIS mixer with a much lower frequency multiplication from a pair of synthesizer.

\section{Conclusion}

A dual-band dual-polarization cryogenic receiver set is proposed for the Black Hole Explorer (BHEX) mission. The primary science instrument, a 240-320 GHz SIS receiver, operating in the DSB mode, will deliver quantum limited sensitivity. A complementary HEMT based receiver covering 80-106 GHz will enhance the signal-to-noise ratio at the correlator output via the Frequency Phase Transfer Technique. 
This pairing of these 2 cryogenic receivers is poised to facilitate the detection of the photon ring of black holes. Strategic trade-offs are being employed to minimize the footprint of the receiver, while not sacrificing the agility and sensitivity of the instrument.

\acknowledgments 

Technical and concept studies for BHEX have been supported by the Smithsonian Astrophysical Observatory, by the Internal Research and Development (IRAD) program at NASA Goddard Space Flight Center, by the University of Arizona, and by the ULVAC-Hayashi Seed Fund from the MIT-Japan Program at MIT International Science and Technology Initiatives (MISTI). We acknowledge financial support from the Brinson Foundation, the Gordon and Betty Moore Foundation (GBMF-10423), and the National Science Foundation (AST-2307887, AST-2107681, AST-1935980, and AST-2034306). This work was supported by the Black Hole Initiative at Harvard University, which is funded by grants from the John Templeton Foundation and the Gordon and Betty Moore Foundation to Harvard University.  BHEX is supported by initial funding from Fred Ehrsam. 

\bibliography{report} 
\bibliographystyle{spiebib} 

\end{document}